\documentclass{bmvc2k}

\usepackage{amsmath, amssymb}
\usepackage{booktabs}
\usepackage{multirow, multicol}
\usepackage{bm}
\usepackage{hyperref}

\title{Out-Of-Distribution Detection for Audio-visual Generalized Zero-Shot Learning: A General Framework}

\addauthor{Liuyuan Wen}{lywen@mail.ustc.edu.cn}{1}
\addauthor{Accepted by BMVC2024}{https://bmvc2024.org/}{1}

\addinstitution{
	School of Physical Sciences\\
	University of Science and Technology of China\\
	Hefei, China
}

\runninghead{Liuyuan Wen}{OOD Detection for AV-GZSL: A General Framework}

\def\etal{\emph{et al}\bmvaOneDot}

\begin{document}

\maketitle

\begin{abstract}
	Generalized Zero-Shot Learning (GZSL) is a challenging task requiring accurate classification of both seen and unseen classes. Within this domain, Audio-visual GZSL emerges as an extremely exciting yet difficult task, given the inclusion of both visual and acoustic features as multi-modal inputs.
	Existing efforts in this field mostly utilize either embedding-based or generative-based methods. However, generative training is difficult and unstable, while embedding-based methods often encounter domain shift problem. Thus, we find it promising to integrate both methods into a unified framework to leverage their advantages while mitigating their respective disadvantages.
	Our study introduces a general framework employing out-of-distribution (OOD) detection, aiming to harness the strengths of both approaches. We first employ generative adversarial networks to synthesize unseen features, enabling the training of an OOD detector alongside classifiers for seen and unseen classes. This detector determines whether a test feature belongs to seen or unseen classes, followed by classification utilizing separate classifiers for each feature type.
	We test our framework on three popular audio-visual datasets and observe a significant improvement comparing to existing state-of-the-art works. Codes can be found in \href{https://github.com/liuyuan-wen/AV-OOD-GZSL}{https://github.com/liuyuan-wen/AV-OOD-GZSL}.
\end{abstract}

\section{Introduction}
\label{sec:intro}

The fusion of acoustic and visual cues in human communication and scene comprehension enjoys widespread recognition, demonstrating notable advancements in various applications such as action recognition \cite{action1, action2}, emotion recognition \cite{emotion1, emotion2}, speech recognition \cite{speech1, speech2} and so on. Despite these advancements, gathering sufficient annotated data for task-specific audio-visual representations remains a daunting challenge. Zero-shot learning (ZSL) arises as a feasible solution, enabling the classification of instances from novel classes by leveraging knowledge acquired from known classes. The more intricate Generalized ZSL (GZSL) further extends this capability to classify test samples from both seen and unseen classes. In this work, we tackle Audio-Visual Generalized Zero-Shot Learning (AV-GZSL) classification task .

Recently, several studies \cite{cjme, avgzsl} have delved into the field of AV-GZSL, innovating methods that project both modalities into a common embedding space and measure distances from class label text embeddings. Mercea \etal \cite{avca} introduced a integrated framework, employing a transformer-based cross-attention mechanism on temporally averaged audio and visual input features. Subsequently, numerous subsequent works have emerged, such as \cite{tcaf, vib, avmst, hyper, mdft}, including a partially generative approach \cite{avfs} which synthesizes unseen features as negative samples in contrastive loss. However, all these approaches are essentially embedding-based and suffer significantly from the domain shift problem \cite{gzsl-review}. Although they employed calibrated stacking \cite{cali} to reduce the bias, it is merely a temporary solution with limited effectiveness.

From our perspective, on one hand, it is crucial to devise a comprehensive and promising strategy to address the domain shift problem. On the other hand, we also acknowledge the difficulty and instability of generative training, and the bias problem inherent in embedding methods. Therefore, we believe that integrating both approaches into a unified framework holds promise, allowing us to leverage the advantages of each method while mitigating their respective disadvantages.

\begin{figure}
	\begin{center}
		\includegraphics[width=1.0\textwidth]{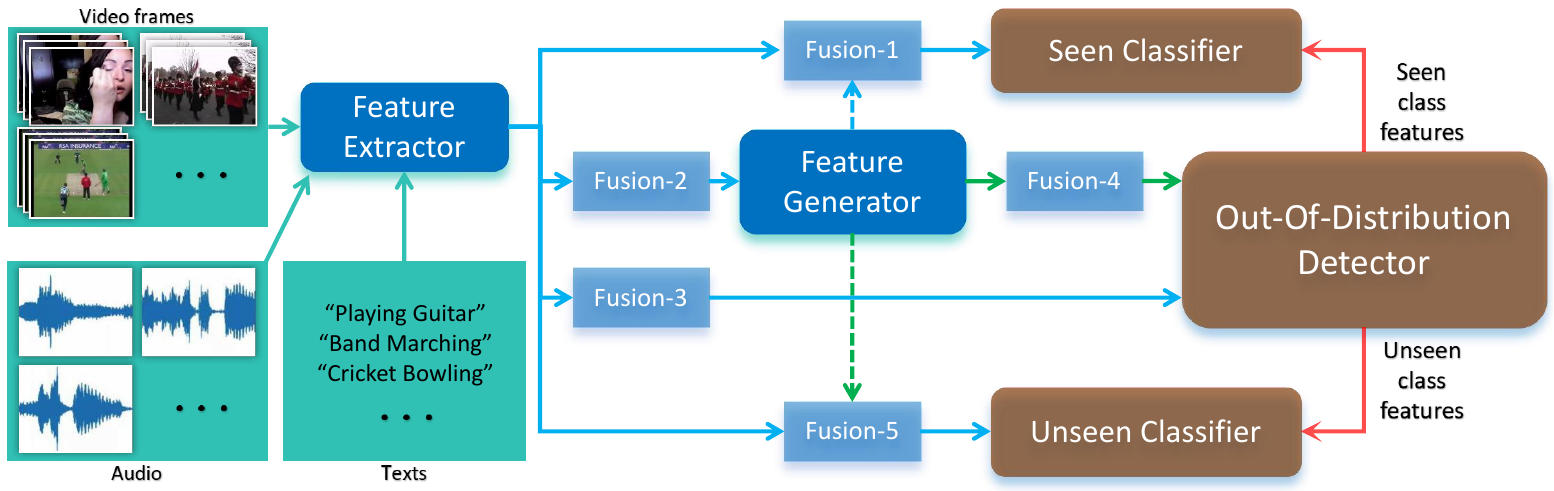}
	\end{center}
	\caption{Our general framework. A generator is used to synthesize features, enabling the training of an OOD detector and separate classifiers for seen and unseen classes. OOD detector distinguishes seen and unseen features for separate classification. Different modality fusion methods (\textit{Fusion 1-5}) can be implemented in different stages. (Blue and green arrows stands for training process using seen and synthesized unseen features respectively. Red arrows represents evaluation process. Dashed lines means ``optional''.)}
	\label{fig:framework}
\end{figure}

In this study, we present a general framework named Audio-Visual Out-Of-Distribution (AV-OOD), as depicted in Figure \ref{fig:framework}. We first employ a generative model to synthesize unseen features, facilitating the training of an OOD detector alongside two separate classifiers for seen and unseen classes. Given the significance of modality fusion in multi-modal learning, different fusion methods (\textit{Fusion 1-5}) can be implemented at various stages to fine-tune model design. During evaluation, the detector assesses whether a test feature belongs to seen or unseen classes, followed by classification using separate classifiers tailored to each feature type. In our experiment, we implement WGAN-GP \cite{wgan, wgan-gp} as the feature generator,  two Multilayer Perceptrons (MLPs) for both the OOD detector and the seen classifier. Lastly, an embedding-based model, adapted from \cite{avca}, serves as the unseen classifier.

Furthermore, the generator can optionally synthesize novel seen or unseen features to aid in classifier training. However, in this work, we refrain from this approach due to our observation that the performance of our generator is not sufficiently satisfactory for synthesizing discriminative trainable samples. Hence, we leave this option for future investigation. This decision also explains why we exclusively utilize the generator for training the OOD detector, as the precise manifold and distribution structure of unseen classes is less critical when only distinguishing between seen and unseen classes.

This framework is designed to be universal, as each component of the model operates relatively independently. Although we will introduce specific models for each part in Section \ref{sec:model}, researchers can enhance overall performance by merely substituting any of them. For instance, models that excel in ZSL, such as spiking neural networks, can replace the unseen classifier, or a superior feature generative network can replace the WGAN-GP used in this work.

\noindent
\textbf{Contributions:}

\noindent
(1) We are the first to introduce OOD detection to the field of AV-GZSL, alongside a general framework integrating both embedding-based and generative-based methods. This integration enables us to effectively harness the advantages of both approaches. 

\noindent
(2) Each component of the framework operates relatively independently and each can be replaced with more effective alternatives, facilitating future researches in terms of structuration and overall performance.

\noindent
(3) We evaluate our framework on three widely-used audio-visual datasets and observe a significant improvement compared to existing state-of-the-art approaches.

\section{Related Works}
\label{sec:rel}

\subsection{Audio-visual Generalized Zero-Shot Learning}
\label{sec:rel:avgzsl}
In the realm of AV-GZSL, embedding-based models are proposed aiming to map video, audio, and text into a shared feature space for comparison. Initially proposed in~\cite{cjme}, Coordinated Joint Multimodal Embedding (CJME) utilized triplet loss for proximity between modal features and class features. Audio-Visual Generalized Zero-shot Learning Network (AVGZSLNet)~\cite{avgzsl} introduced a module reconstructing text features from visual and audio inputs. Audio-Visual Cross-modal Attention (AVCA)~\cite{avca} leveraged cross-modal attention mechanisms, effectively integrates information from both modalities. Temporal Cross-attention Framework (TCAF)~\cite{tcaf} proposed a temporal cross-attention framework to enhance cross-attention across modalities and time. Audio-Visual Feature Synthesis (AVFS)~\cite{avfs} simulated unseen features with generative models, combining contrastive and discriminative losses. Variational Information Bottleneck for AV-GZSL (VIB-GZSL)~\cite{vib} is a method based on variational information bottleneck. Audio-Visual Modality-fusion Spiking Transformer (AVMST)~\cite{avmst} and Motion-Decoupled Spiking
Transformer (MDFT)~\cite{mdft} utilized spiking neural networks to process highly sparsity event data efficiently. Hyper-multiple~\cite{hyper} used a hyperbolic transformation to achieve curvature-aware geometric learning.

All works above are essentially embedding-based methods including AVFS which only did an shallow addition and still suffered from bias problem. Our proposed framework is a fundamental solution for bias problem by implementing OOD detection, which integrates both embedding-based and generative-based methods.

\subsection{Out-Of-Distribution Detection}
Out-of-distribution (OOD) detection serves as a crucial method in GZSL, focusing on distinguishing between seen and unseen classes during inference to prevent the model from biased predictions towards the known categories. Socher \etal \cite{ood1} pioneered the introduction of a binary novelty random variable for novelty detection. Atzmon \etal \cite{ood2} proposed an adaptive confidence smoothing method, serving as a gating model for OOD detection. Subsequently, a variety of methods emerged, including entropy-based \cite{cewgan-od}, probabilistic-based \cite{ood3}, and boundary-based \cite{ood4}, among others.

In this study, we primarily employ entropy-based methods \cite{cewgan-od} and integrate them into AV-GZSL alongside more expert classifiers. We evaluate two types of entropy losses and conduct a comparative analysis against the traditional calibrated stacking for the first time.

\section{Model Architecture}
\label{sec:model}

\begin{figure}
	\begin{center}
		\includegraphics[width=1.0\textwidth]{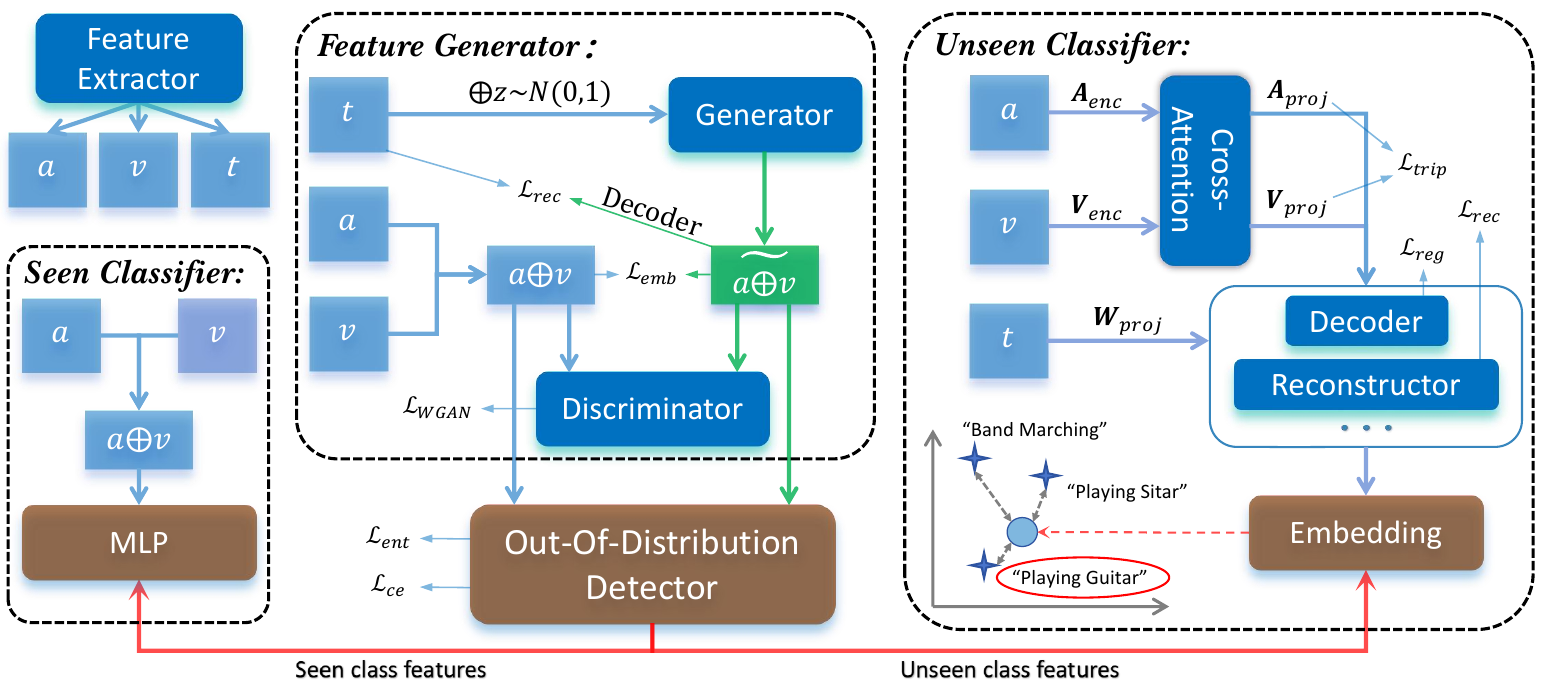}
	\end{center}
	\caption{We employ MLPs as the seen classifier and OOD detector, an embedding-based model as the unseen classifier, and WGAN-GP for feature generation. Extracted features $a$, $v$, $t$ stands for audio, visual, text respectively. Fusion methods involve simple concatenation $\oplus$ except for \textit{Fusion-5} in the unseen classifier, which utilizes cross-attention. The synthesized unseen features $\widetilde{a\oplus v}$ along with seen ones $a\oplus v$ are input for training of the OOD Detector, which distinguishes seen and unseen features for separate classification during test stage. (Blue and green arrows depict the training process using real seen and synthesized unseen features respectively. Red arrows represents evaluation process.)}
	\label{fig:structure}
\end{figure}

In this section, we will introduce each component of the framework utilized in this study as Figure \ref{fig:structure} shows. In general, we employ a conditional WGAN-GP \cite{wgan, wgan-gp} as the feature generator, an MLP equipped with entropy-based loss as the OOD detector \cite{cewgan-od}, another MLP serving as the seen classifier, and an embedding-based model adapted from AVCA \cite{avca} as the unseen classifier.

\subsection{Feature Generator}
We opt for a conditional WGAN-GP with an additional loss as our feature generator, inspired by its recent success in video classification \cite{cewgan-od}. On top of that, AV-GZSL requires an additional step of modality fusion, where we simply concatenate audio and visual features as $x=a\oplus v$. And the synthesized features are represented as $\tilde{x}=\widetilde{a\oplus v}$. The training process involves three distinct loss components: $\mathcal{L}_{\text {WGAN}}$, $\mathcal{L}_{rec}$ and $\mathcal{L}_{emb}$.

\noindent
\textbf{Conditional WGAN-GP loss.}
A generator $G$ and a discriminator $D$ are employed in GAN. We denote the generated feature as $\tilde{x}=G(z,t)$ where $z\sim N(0,1)$, the convex combination of $x$ and $\tilde{x}$ as $\hat{x}$, and the penalty coefficient as $\lambda$. Thus, conditioned on the textual embedding $t$, the expression is:
\begin{equation}
	\mathcal{L}_{\text{WGAN}}= \mathbb{E}[D(x,t)]-\mathbb{E}[D(\widetilde{x}, t)]- \lambda \mathbb{E}\left[\left(\left\|\nabla_{\hat{x}} D(\hat{x}, t)\right\|_2-1\right)^2\right].
\end{equation}

\noindent
\textbf{Reconstruction loss.}
A decoder is utilized to reconstruct the class embedding $t$ from the synthesized $\tilde{x}$ for generated features to be discriminative. Let $\mathcal{O}^{dec}$ represent the outputs of the decoder, and $\mathbf{MSE}$ represent the Mean Square Error metric, then $\mathcal{L}_{rec}=\mathbf{MSE} (\mathcal{O}^{dec}, t)$.

\noindent
\textbf{Embedding loss.} We organize the real and synthesized features as matched (same classes) and unmatched (different classes) pairs. Then, a Cosine Embedding loss $\mathbf{CE}$ is employed to minimize and maximize the distance between the matched and unmatched features, respectively. Thus, $\mathcal{L}_{emb}=\mathbf{CE}(x, \tilde{x})$.

Thereby the total loss for feature generator is: 
\begin{equation}
	\mathcal{L}_{gen} = \min_G \max_D \mathcal{L}_{WGAN}+\alpha \mathcal{L}_{rec} + \beta \mathcal{L}_{emb} \label{lgen},
\end{equation}
where $\alpha$ and $\beta$ are weight coefficients.

\subsection{Out-Of-Distribution Detector}
\label{sec:ood}
Our OOD detector takes real seen and synthesized unseen features as positive and negative inputs, respectively. It outputs the probability of samples belonging to seen or unseen classes. In practice, we set a specific threshold, where outputs exceeding or falling below are categorized as seen or unseen accordingly.

We evaluate two types of OOD detectors, binary and entropy classifiers (See Section \ref{sec:ablation}), both implemented as 3-layer MLPs denoted as $O_{bin}$ and $O_{ent}$.

$O_{bin}$ is a straightforward classifier using Binary Cross Entropy loss
\begin{equation}
	\mathcal{L}_{bin}=\mathbf{BCE}\left( O_{bin}(x),1 \right) + \mathbf{BCE}\left( O_{bin}(\tilde{x}),0 \right).
\end{equation}
However, $O_{bin}$ proves inadequate due to the complex boundaries between seen and unseen classes (See Section \ref{sec:ab-bias}).
Therefore, we adopt an information entropy-based classifier with loss 
\begin{equation}
	\mathcal{L}_{ent}=\mathbf{H}\left( O_{ent}(x) \right) - \mathbf{H}\left( O_{ent}(\tilde{x}) \right),
\end{equation}
where $\mathbf{H}(p)=-\sum_i p_i \log (p_i)$ represents the entropy of $p$. Entropy effectively quantifies the uncertainty of features; the lower a feature's entropy, the more likely it belongs to seen classes. Additionally, we include a Cross Entropy loss to accelerate convergence:
\begin{equation}
	\mathcal{L}_{ce} = \mathbf{CE} (O_{ent}(x),y(x)),
\end{equation}
where $y(x)$ is the corresponding label of sample $x$. Thus, the total loss for the OOD detector becomes $\mathcal{L}_{ood} = \mathcal{L}_{ent} + \mathcal{L}_{ce}$.

\subsection{Seen Classifier}
We adopt a 3-layer MLP for our seen classifier, finding it more stable and efficient compared to overly complex model structures (See Section \ref{sec:ab-cls}). It takes $x=a\oplus v$ as input and employs Cross Entropy loss $\mathcal{L}_{sc} = \mathbf{CE} \left( x,y(x) \right)$, $y(x)$ is the corresponding label of sample $x$.

\subsection{Unseen Classifier}
We enhance the AVCA \cite{avca} framework by introducing an additional negative loss component and take it as our unseen classifier. Instead of simple concatenation, we employ a transformer-based cross-attention for modality fusion.

As shown on the right side of Figure \ref{fig:structure}, in each training session, features $a$ and $v$ undergo a cross-attention block after being embedded into the same dimension as $t$ by encoders $\bm{A_{enc}}$ and $\bm{V_{enc}}$. This block facilitates the exchange of semantics among different modalities, capitalizing on shared cross-modal information.
Subsequently, projectors $\bm{V_{proj}}$, $\bm{A_{proj}}$, and $\bm{W_{proj}}$ project all modalities of samples into a common embedding space. Following this, the remaining procedures, including decoding, reconstruction, and backward process, are conducted for loss calculation.
During evaluation, all three modalities are re-projected into the same joint embedding space as described above for classification. The closest textual label embedding of the current sample is then selected as the classified target.

To ensure effective feature representation learning, the loss function $\mathcal{L}_{uc}$ consists of three components: triplet loss $\mathcal{L}_{trip}$, reconstruction loss $\mathcal{L}_{rec}$, and regularization loss $\mathcal{L}_{reg}$. While a contrastive learning strategy typically randomly selects a negative sample from a different class for every feature in the baseline, it's important to note that previous studies often apply these losses solely for positive samples, neglecting losses from negative samples. To promote more robust training, we treat them equally and compute all three types of losses for both positive and negative samples.

\noindent
\textbf{Triplet loss}. Four types of triplet loss functions cooperate to align contrastive audio, visual, and textual label embeddings:
\begin{align}
	\begin{split}
		\mathcal{L}_{trip}^{\pm} = 
		& \mathbf{T} \left( \mathcal{O}_{av}^{proj,\pm}, \mathcal{O}_{t}^{proj,\pm}, \mathcal{O}_{av}^{proj,\mp} \right)
		+ \mathbf{T} \left( \mathcal{O}_{t}^{proj,\pm}, \mathcal{O}_{av}^{proj,\pm}, \mathcal{O}_{t}^{proj,\mp} \right) \\ 
		+ & \mathbf{T} \left( \mathcal{O}_{t}^{proj,\pm}, \mathcal{O}_{av}^{proj,\pm}, \mathcal{O}_{av}^{proj,\mp} \right)
		+ \mathbf{T} \left( \mathcal{O}_{av}^{proj,\pm}, \mathcal{O}_{t}^{proj,\pm}, \mathcal{O}_{t}^{proj,\mp} \right),
	\end{split}
\end{align}
where $\mathbf{T}(\cdot)$ represents the triplet loss function $\mathbf{T}(x, y, z) = \max \{\left\lVert x - y \right\rVert_2 - \left\lVert x - z \right\rVert_2 + \mu, 0\}$. Here, $x$, $y$, and $z$ are anchor embeddings, and $\mu$ denotes the margin. $\mathcal{O}_{m}^{proj,\pm}$ represents the output from the projector in the joint space, where $m \in \{ a, v, t\}$ denotes three modalities. Superscripts $\{+, -\}$ represent positive and negative samples, respectively.
These four types of triplet loss functions cooperate to
align contrastive audio, visual, and textual label embeddings.

\noindent
\textbf{Reconstruction loss}. A decoder reconstructs and aligns the initial and final states of textual features. We define 
\begin{equation}
	\mathcal{L}_{rec}^{\pm} = \sum_{m \in \{a,v,t\}} \mathbf{MSE} \left( \mathcal{O}_{m}^{dec,\pm}, t^{\pm} \right),
\end{equation}
where $\mathcal{O}_{m}^{dec,\pm}$ denotes the output tensor from decoder operators, considered as the final state, and $t^{\pm}$ represents the input textual features, considered as the initial state.

\noindent
\textbf{Regularization loss}. The regularization loss encourages the alignment of audio and visual embeddings with text embeddings while preserving information from their respective input modalities. It is expressed as 
\begin{equation}
	\mathcal{L}_{reg}^{\pm} = \sum_{m \in \{a,v\}} \mathbf{MSE} \left( \mathcal{O}_{m}^{rec,\pm}, \mathcal{O}_{m}^{enc,\pm} \right), 
\end{equation}
where $\mathcal{O}_{m}^{rec,\pm}$ and $\mathcal{O}_{m}^{enc,\pm}$ represent output samples from reconstructors and encoders, respectively.

Thus, the total loss of the unseen classifier is formulated as $\mathcal{L}_{uc} = \mathcal{L}^{+} + \mathcal{L}^{-}$, where $\mathcal{L}^{\pm} = \mathcal{L}_{trip}^{\pm} + \mathcal{L}_{rec}^{\pm} + \mathcal{L}_{reg}^{\pm}$.

\section{Experiments}
\subsection{Experimental Setup}
We adopt AVCA \cite{avca} as our baseline, along with a commonly used benchmark proposed by them on datasets curated from VGGSound~\cite{vgg}, UCF101~\cite{ucf}, and ActivityNet~\cite{activity}. Notably, AVCA employs a two-stage training and evaluation protocol where epochs and calibrated stacking of the second stage are determined by the best performance on the validation set in the first stage. Since we have replaced calibrated stacking with OOD detection, we only run the second stage with all epochs fixed, which is equivalent.


For evaluation on the test dataset, we report mean class accuracy for both the Seen ($\mathbf{S}$) and Unseen ($\mathbf{U}$) subsets, with the Harmonic Mean $\mathbf{HM} = \frac{2\mathbf{U}\mathbf{S}}{\mathbf{U}+\mathbf{S}}$. The Zero-Shot Learning ($\mathbf{ZSL}$) performances are determined solely by evaluating on the test unseen subset.

\subsection{Implementation Details}
\noindent
\textbf{Feature Extractor:}
Following \cite{avca}, we extract audio features $a$ and visual features $v$, and average them second-wise using self-supervised SeLaVi~\cite{selavi} framework pretrained on VGGSound\cite{vgg}. Textual label embeddings $t$ are acquired through word2vec network pretrained on Wikipedia~\cite{word2vec}.

\noindent
\textbf{Feature Generator:}
We employ WGAN-GP \cite{wgan, wgan-gp} with the loss function in Equation \ref{lgen}. In our experiments, we set $\alpha$ to 0.1 and $\beta$ to 0.01. Training is conducted over 5 epochs with a learning rate of 0.0001, using the Adam optimizer.

\noindent
\textbf{Out-Of-Distribution Detector:}
Our detector is a 3-layer MLP with hidden layers of dimensions 512 and 128. For training, we generate 50, 50, and 1000 unseen samples from VGGSound, UCF, and ActivityNet, respectively. Training spans 80 epochs with learning rates and batch sizes of 0.001/6900 for VGGSound, 0.009/16 for UCF, and 0.005/64 for ActivityNet.

\noindent
\textbf{Seen Classifier:}
The seen classifier is also a 3-layer MLP, featuring hidden layers of dimensions 512 and 256. Training is conducted over 200 epochs with learning rates and batch sizes of 0.008/1024 for VGGSound, 0.0006/32 for UCF, and 0.008/128 for ActivityNet.

\noindent
\textbf{Unseen Classifier:}
We utilize triplet loss with a margin $\mu$ of 1. Dropout rates for the encoder/projector/decoder are set at 0.3/0.1/0.2 for VGGSound, 0.5/0.4/0.4 for UCF, and 0.2/0.3/0.2 for ActivityNet. Training spans 50 epochs with learning rates and batch sizes of 0.0005/256 for VGGSound, 0.0024/112 for UCF, and 0.0005/256 for ActivityNet.

\subsection{Results Analysis}
\noindent
\textbf{Compared methods.}
We compare our AV-OOD with current state-of-the-art AV-GZSL models as introduced in Section~\ref{sec:rel:avgzsl}, with methods range from embedding-based ones to generative-based ones. The results are presented in Table~\ref{tab:score}.

\noindent
\textbf{Comparison of performances.}
From Table \ref{tab:score}, it is evident that AV-OOD stands out as the top performer across all three datasets. Specifically, AV-OOD achieves a remarkable $\mathbf{HM}$ score of 11.16\% on the VGGSound dataset, marking a substantial 77\% enhancement over the baseline AVCA score of 6.31\%. Notably, AV-OOD also exhibits outstanding performance on UCF and ActivityNet, with $\mathbf{HM}$ scores of 37.89\% and 14.38\% respectively, surpassing all competing models. Moreover, the \textbf{S} scores of AV-OOD demonstrate notable stability, particularly evident on VGGSound and ActivityNet, recording 21.25\% and 31.9\% respectively. This commendable performance can be attributed not only to the implementation of a more efficient MLP for replacing the seen classifier but also to the replacement of calibrated stacking, which significantly curtails the model's ability on seen classes.
However, it is noteworthy that while AV-OOD excels in $\mathbf{HM}$ scores, its $\mathbf{ZSL}$ scores do not achieve the same level of prominence. This underscores the primary objective of our framework, which prioritizes enhancing the model's discrimination between seen and unseen classes rather than focusing solely on ZSL skills. Nonetheless, despite not reaching the top position in $\mathbf{ZSL}$, AV-OOD remains notable for its superiority over the baseline AVCA~\cite{avca} across all metrics. This can be attributed to our incorporation of losses from negative samples, which evidently enhance the model's robustness and learning efficiency.

\begin{table*}[t]
	\begin{center}
		\setlength{\tabcolsep}{3pt} 
		\caption{Comparison of AV-GZSL performances} 
		\label{tab:score}
		\scalebox{0.82}{
			\begin{tabular}{clcccclcccclcccc}
				\toprule
				\multirow{2.5}{*}{Model} && \multicolumn{4}{c}{VGGSound-GZSL} && \multicolumn{4}{c}{UCF-GZSL} && \multicolumn{4}{c}{ActivityNet-GZSL} \\
				\cmidrule(lr){3-6} \cmidrule(lr){8-11} \cmidrule(lr){13-16}
				&& $\mathbf{S}$ & $\mathbf{U}$ & $\mathbf{HM}$ & $\mathbf{ZSL}$ && $\mathbf{S}$ & $\mathbf{U}$ & $\mathbf{HM}$ & $\mathbf{ZSL}$ && $\mathbf{S}$ & $\mathbf{U}$ & $\mathbf{HM}$ & $\mathbf{ZSL}$ \\
				\midrule
				CJME\cite{cjme} && 8.69 & 4.78 & 6.17 & 5.16 && 26.04 & 8.21 & 12.48 & 8.29 && 5.55 & 4.75 & 5.12 & 5.84 \\
				AVGZSLNet\cite{avgzsl} && 18.15 & 3.48 & 5.83 & 5.28 && 52.52 & 10.90 & 18.05 & 13.65 && 8.93 & 5.04 & 6.44 & 5.40 \\
				AVCA\cite{avca} && 14.90 & 4.00 & 6.31 & 6.00 && 51.53 & 18.43 & 27.15 & 20.01 && 24.86 & 8.02 & 12.13 & 9.13 \\
				TCAF\cite{tcaf} && 9.64 & 5.91 & 7.33 & 6.06 && 58.60 & 21.74 & 31.72 & 24.81 && 18.70 & 7.50 & 10.71 & 7.91 \\
				VIB-GZSL\cite{vib} && $\underline{18.42}$ & 6.00 & 9.05 & 6.41 && \textbf{90.35} & 21.41 & $\underline{34.62}$ & 22.49 && 22.12 & 8.94 & 12.73 & 9.29 \\
				AVFS\cite{avfs} && 15.20 & 5.13 & 7.67 & 6.20 && 54.87 & 16.49 & 25.36 & 22.37 && $\underline{29.00}$ & 9.13 & $\underline{13.89}$ & 11.18 \\
				AVMST\cite{avmst} && 14.14 & 5.28 & 7.68 & 6.61 && 44.08 & 22.63 & 29.91 & 28.19 && 17.75 & $\underline{9.90}$ & 12.71 & 10.37 \\
				MDFT\cite{mdft} && 16.14 & 5.97 & 8.72 & 7.13 && 48.79 & $\underline{23.11}$ & 31.36 & \textbf{31.53} && 18.32 & \textbf{10.55} & 13.39 & \textbf{12.55} \\
				Hyper-multiple\cite{hyper} && 15.02 & $\underline{6.75}$ & $\underline{9.32}$ & $\underline{7.97}$ && 63.08 & 19.10 & 29.32 & 22.24 && 23.38 & 8.67 & 12.65 & 9.50 \\
				\textbf{AV-OOD (ours)} && \textbf{21.25} & \textbf{7.57} & \textbf{11.16} & \textbf{8.78} && $\underline{65.78}$ & \textbf{26.61} & \textbf{37.89} & $\underline{28.21}$ && \textbf{31.9} & 9.28 & \textbf{14.38} & $\underline{11.49}$ \\
				\bottomrule
			\end{tabular}
		}
	\end{center}
\end{table*}

\subsection{Ablation Study}
\label{sec:ablation}

\subsubsection{Effect of Different Bias Reduction Methods}
\label{sec:ab-bias}

\begin{figure}
	\begin{center}
		\includegraphics[width=1.0\textwidth]{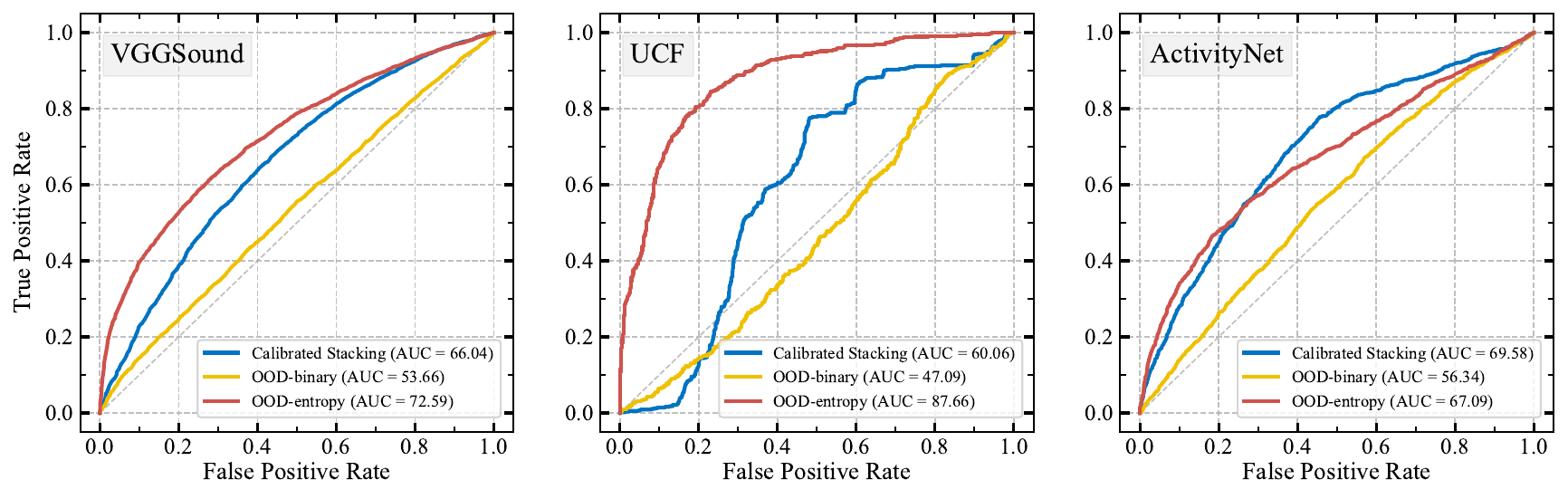}
	\end{center}
	\caption{Comparison between ROC curves of three bias reduction methods on all three datasets. Obviously, our proposed ``OOD-entropy'' is the most stable one and gets the best \textbf{AUC} on two of the three datasets.}
	\label{fig:roc}
\end{figure}

\begin{table*}[t]
	\begin{center}
		\setlength{\tabcolsep}{3pt} 
		\caption{Comparison of different bias reduction methods} 
		\label{tab:roc}
		\scalebox{0.85}{
			\begin{tabular}{clccclccclccc}
				\toprule
				\multirow{2.5}{*}{Method} && \multicolumn{3}{c}{VGGSound} && \multicolumn{3}{c}{UCF} && \multicolumn{3}{c}{ActivityNet} \\
				\cmidrule(lr){3-5} \cmidrule(lr){7-9} \cmidrule(lr){11-13}
				&& $\mathbf{AUC}$ & $\mathbf{FPR}$ & $\mathbf{HM}$ && $\mathbf{AUC}$ & $\mathbf{FPR}$ & $\mathbf{HM}$ && $\mathbf{AUC}$ & $\mathbf{FPR}$ & $\mathbf{HM}$ \\
				\midrule
				Calibrated Stacking && 66.04 & 36.43 & 9.32 && 60.06 & 39.78 & 27.57 && \textbf{69.58} & \textbf{30.82} & \textbf{12.04} \\
				OOD-binary && 53.66 & 54.9 & 7.82 && 47.09 & 63.38 & 17.9 && 56.34 & 50.94 & 10.44 \\
				\textbf{OOD-entropy (ours)} && \textbf{72.59} & \textbf{26.64} & \textbf{10.11} && \textbf{87.66} & \textbf{8.68} & \textbf{37.03} && 67.09 & 33.87 & 11.56 \\
				\bottomrule
			\end{tabular}
		}
	\end{center}
\end{table*}

As introduced in Section \ref{sec:intro}, domain shift problem arises due to discrepancies between the seen and unseen data distributions across different semantic domains. Previous works on AV-GZSL all adopted Calibrated Stacking \cite{cali} as the bias reduction method, which we replaced with OOD detection. We tested three bias reduction methods while preserving the performances of seen and unseen classifiers: \textit{Calibrated Stacking}, \textit{OOD-binary}, and \textit{OOD-entropy}, to compare their capability of distinguishing between seen and unseen classes. Seen classes are regarded as positive and in-distribution, while unseen classes are considered negative and out-of-distribution.

We plot Receiver Operating Characteristic (ROC) curves with True Positive Rate (\textbf{TPR}) and False Positive Rate (\textbf{FPR}) by sweeping over classification thresholds. Figure \ref{fig:roc} displays all the ROC curves along with their Area Under Curve (\textbf{AUC}). \textbf{FPR} and \textbf{HM} in Table \ref{tab:roc} correspond to the threshold that yields a 60\% \textbf{TPR} for detecting in-distribution samples. 
We can observe that \textit{OOD-entropy} is the most stable method and achieves the best AUC on two of the three datasets. Taking UCF as an example, \textit{OOD-entropy} obtains an \textbf{AUC} of 87.66\%, significantly surpassing 60.06\% of \textit{Calibrated Stacking} and 47.09\% of \textit{OOD-binary}. 
Meanwhile, \textit{OOD-entropy} achieves the lowest \textbf{FPR} of 8.68\%, compared to 39.78\% of \textit{Calibrated Stacking} and 63.38\% of \textit{OOD-binary}. This contributes to the highest \textbf{HM} of \textit{OOD-entropy}, reaching 37.03\%. 

We observed that \textit{Calibrated Stacking} and \textit{OOD-binary} fall below the random guess line in certain areas. This is because UCF contains too few samples for the generator and classifiers to learn a comprehensive representation, which can be well supplemented using \textit{OOD-entropy}.
Moreover, \textit{OOD-binary} performs the worst across all three datasets, confirming our judgment in Section \ref{sec:ood} that a binary classifier is too simplified to learn the complex boundaries between seen and synthesized unseen samples whose precision is also limited by the generator. 

Lastly, although \textit{Calibrated Stacking} outperforms \textit{OOD-entropy} on ActivityNet, its performance declines quickly when \textbf{TPR} (or \textbf{FPR}) is relatively low, as seen in Figure \ref{fig:roc}, where the best \textbf{HM} often appears. Thus, \textit{OOD-entropy} still holds the first place with the final best \textbf{HM} of 14.38\% (See Table \ref{tab:score}).

\subsubsection{Effect of Different Classifier Models}
\label{sec:ab-cls}
We conducted tests using different choices of seen and unseen classifiers: \textit{MLP} or \textit{Embedding} methods, and the results are presented in Table \ref{tab:classifier}. $\mathbf{SC_{acc}}$ and $\mathbf{UC_{acc}}$ represent the accuracy of the seen and unseen classifiers, respectively. \textbf{HM} is obtained with the assistance of OOD-entropy.

It can be concluded that \textit{MLP} is the better choice for seen classes, while \textit{Embedding} is better for unseen classes. For example, on VGGSound, \textit{MLP} achieves a $\mathbf{SC_{acc}}$ of 58.27\% compared to \textit{Embedding}'s 49.42\%. However, the $\mathbf{UC_{acc}}$ is only 3.39\% compared to \textit{Embedding}'s 8.78\%. The same trend applies to the other two datasets.

This difference stems from two main aspects. Firstly, our embedding model is originally designed for ZSL and may be too complex for familiar class classification with the risk of overfitting, whereas a simple \textit{MLP} can perform this task proficiently. Secondly, the performance of $\mathbf{UC_{acc}}$ of \textit{MLP} is limited because, unlike \textit{Embedding}, it can only learn from synthesized unseen features due to its fixed output dimension, without including the more crucial real seen features. However, there is considerable room for improvement in terms of both $\mathbf{SC_{acc}}$ and $\mathbf{UC_{acc}}$ across all datasets. This deficiency could potentially be addressed by employing more powerful specialized classification models in future works.

\begin{table*}[t]
	\begin{center}
		\setlength{\tabcolsep}{3pt} 
		\caption{Comparison of different classifier models} 
		\label{tab:classifier}
		\scalebox{0.8}{
			\begin{tabular}{clccclccclccc}
				\toprule
				\multirow{2.5}{*}{Classifiers (Seen \& Unseen)} && \multicolumn{3}{c}{VGGSound} && \multicolumn{3}{c}{UCF} && \multicolumn{3}{c}{ActivityNet} \\
				\cmidrule(lr){3-5} \cmidrule(lr){7-9} \cmidrule(lr){11-13}
				&& $\mathbf{SC_{acc}}$ & $\mathbf{UC_{acc}}$ & $\mathbf{HM}$ && $\mathbf{SC_{acc}}$ & $\mathbf{UC_{acc}}$ & $\mathbf{HM}$ && $\mathbf{SC_{acc}}$ & $\mathbf{UC_{acc}}$ & $\mathbf{HM}$ \\
				\midrule
				MLP \& MLP && \textbf{58.27} & 3.39 & 5.51 && \textbf{96.12} & 17.7 & 26.86 && \textbf{62.33} & 3.71 & 5.81 \\
				Embedding \& Embedding && 49.42 & \textbf{8.78} & 10.97 && 64.97 & \textbf{28.21} & 30.26 && 33.99 & \textbf{11.49} & 12.14 \\
				\textbf{MLP \& Embedding (ours)} && \textbf{58.27} & \textbf{8.78} & \textbf{11.16} && \textbf{96.12} & \textbf{28.21} & \textbf{37.89} && \textbf{62.33} & \textbf{11.49} & \textbf{14.38} \\
				\bottomrule
			\end{tabular}
		}
	\end{center}
\end{table*}

\section{Conclusion}
In this study, we introduce out-of-distribution (OOD) detection into the realm of Audio-Visual Generalized Zero-Shot Learning (AV-GZSL) for the first time. We establish a comprehensive framework that forthcoming researches can adopt to achieve more structural, organized and promising results. We first employ generative adversarial networks to synthesize unseen features, thereby enabling the training of an OOD detector alongside classifiers for both seen and unseen classes. This detector determines whether a test feature belongs to seen or unseen classes, followed by classification utilizing separate classifiers for each feature type. 

Our work demonstrates superiority in two key aspects. Firstly, we integrate both embedding-based and generative-based methods, effectively leveraging the advantages of each approach. Secondly, each component of our framework operates relatively independently and can be substituted with more effective alternatives. This facilitates future exploration into structured models and enhances overall performances.

\bibliography{avood}
\end{document}


\maketitle

In this supplementary material, we present a detailed introduction to the datasets used in Section \ref{sec:exp}. Furthermore, additional ablation studies on the negative loss terms and threshold tuning, along with t-SNE visualizations, are unveiled in Section \ref{sec:ab} and Section \ref{sec:tsne}, respectively.

\section{Datasets}
\label{sec:exp}

In this study, we conduct experiments on three widely used datasets: VGGSound \cite{vgg}, UCF \cite{ucf}, and ActivityNet \cite{activity}.

\noindent
\textbf{VGGSound:} 
The VGGSound dataset is a large-scale audio-visual dataset consisting of short clips of audio sounds extracted from videos uploaded to YouTube. It is designed to be a comprehensive resource for studying the correspondence between audio and visual elements in diverse and challenging acoustic environments. We selected 276 classes that are clearly labeled in our experiment, resulting in 93,752 videos.

\noindent
\textbf{UCF:} 
The UCF101 dataset is widely recognized for action recognition in videos. As an extension of the UCF50 dataset, it is notable for its diversity and complexity, making it one of the most challenging datasets for video-based human action recognition. We included only the 51 classes that contain audio information, resulting in 6,816 videos.

\noindent
\textbf{ActivityNet:} 
The ActivityNet dataset is a large-scale video benchmark designed for human activity understanding. It contains 200 different types of activities across 20,348 video clips collected from YouTube. This dataset is significant for its size and diversity, making it a challenging and comprehensive benchmark for temporal activity detection.

%
%
%

\section{Additional Ablation Study}
\label{sec:ab}
\subsection{Effects of Negative Loss Terms of Unseen Classifier}
\begin{table*}[t]
	\begin{center}
		\setlength{\tabcolsep}{3pt} 
		\caption{Effects of negative loss terms of unseen classifier} 
		\label{tab:loss}
		\scalebox{0.85}{
			\begin{tabular}{clcclcclcc}
				\toprule
				\multirow{2.5}{*}{Loss} && \multicolumn{2}{c}{VGGSound} && \multicolumn{2}{c}{UCF} && \multicolumn{2}{c}{ActivityNet} \\
				\cmidrule(lr){3-4} \cmidrule(lr){6-7} \cmidrule(lr){9-10}
				&& $\mathbf{UC_{acc}}$ & $\mathbf{HM}$ && $\mathbf{UC_{acc}}$ & $\mathbf{HM}$ && $\mathbf{UC_{acc}}$ & $\mathbf{HM}$ \\
				\midrule
				$\mathcal{L}^+$ && 7.83 & 9.92 && 20.69 & 29.86 && 9.9 & 12.83 \\
				$\mathcal{L}^+ + \mathcal{L}_{trip}^-$ && 8.35 & 10.36 && 25.31 & 34.9 && 9.81 & 12.9 \\
				$\mathcal{L}^+ + \mathcal{L}_{rec}^-$ && 7.36 & 9.62 && 18.77 & 27.19 && 9.42 & 12.72 \\
				$\mathcal{L}^+ + \mathcal{L}_{reg}^-$ && 7.33 & 9.43 && 21.81 & 30.99 && 8.88 & 11.97 \\
				$\mathcal{L}^+ + \mathcal{L}^-$(ours) && \textbf{8.78} & \textbf{11.16} && \textbf{28.21} & \textbf{37.89} && \textbf{11.49} & \textbf{14.38} \\
				\bottomrule
			\end{tabular}
		}
	\end{center}
\end{table*}
As explained in Section 3.4, prior research often focuses solely on losses associated with positive samples \cite{vib, avfs, avmst, hyper, mdft}, while neglecting losses from negative samples. To foster more robust training, we accord them equal importance and compute all three types of losses for both positive and negative samples. We present an ablation study of different loss terms in Table \ref{tab:loss}. $\mathbf{UC_{acc}}$ denotes the accuracy of the unseen classifier, while $\mathbf{HM}$ scores are provided with other parts of the model weights fixed for a fair comparison.
It is evident that the inclusion of $\mathcal{L}_{trip}^-$ alone can remarkably enhance scores. Taking VGGSound as an example, $\mathcal{L}^+ + \mathcal{L}_{trip}^-$ yields 8.35\% and 10.36\% for $\mathbf{UC_{acc}}$ and $\mathbf{HM}$, respectively, surpassing $\mathcal{L}^+$ alone with scores of 7.83\% and 9.92\%. On the other hand, while the addition of either $\mathcal{L}_{rec}^-$ or $\mathcal{L}_{reg}^-$ alone does not seem particularly effective, the combination of all three negative loss terms performs admirably. Indeed, $\mathcal{L}^+ + \mathcal{L}^-$ outperforms $\mathcal{L}^+$ in all aspects, achieving the highest scores of 8.78\% and 11.16\% on VGGSound.

\subsection{Parameter Tuning of Thresholds of OOD-entropy}
\begin{figure}[t]
	\begin{center}
		\includegraphics[width=1.0\textwidth]{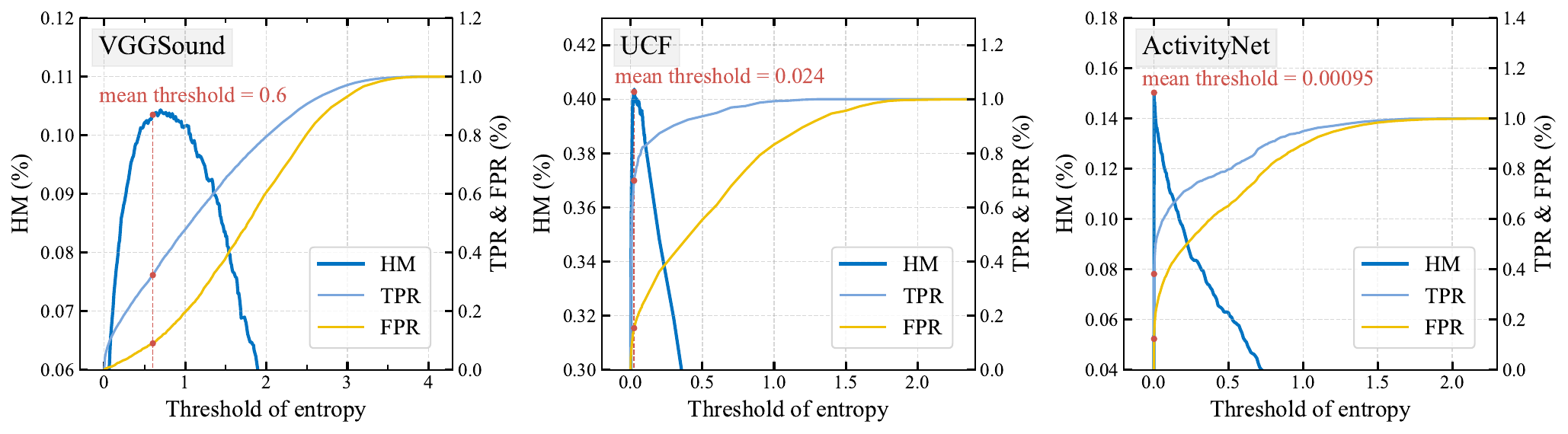}
	\end{center}
	\caption{Effects of OOD-entropy thresholds on $\mathbf{HM}$, $\mathbf{TPR}$, and $\mathbf{FPR}$ across three datasets. In each figure, the left vertical axes correspond to $\mathbf{HM}$ lines, while the right axes correspond to $\mathbf{TPR}$ and $\mathbf{FPR}$ lines. Red dots and words represent the average entropy of seen classes from the training data, which evidently aligns closely with the value maximizing $\mathbf{HM}$.}
	\label{fig:thresholds}
\end{figure}
All three bias reduction methods explored in Section 4.3.1 require parameter tuning of thresholds, outputs exceeding or falling below which are categorized as seen or unseen accordingly. 
Figure \ref{fig:thresholds} illustrates the curves of $\mathbf{HM}$, $\mathbf{TPR}$, and $\mathbf{FPR}$ as they change with entropy thresholds.
Unlike \textit{Calibrated Stacking} and \textit{OOD-binary}, whose thresholds necessitate tuning through iterative processes such as ``for'' loops, the thresholds of \textit{OOD-entropy} can be readily determined using the average entropy of seen classes from the training data (indicated by red dots and the descriptor \textit{mean threshold} in Figure \ref{fig:thresholds}). It's evident that all \textit{mean threshold}s closely align with the value that maximizes $\mathbf{HM}$. Additionally, we observe significant variation in thresholds corresponding to the points with the highest $\mathbf{HM}$ across different datasets, namely 0.6, 0.024, and 0.00095 for VGGSound, UCF, and ActivityNet respectively. This variability underscores the inefficiency of tuning entropy thresholds through iterative loops.

\section{Qualitative Results}
\begin{figure}[t]
	\begin{center}
		\includegraphics[width=1.0\textwidth]{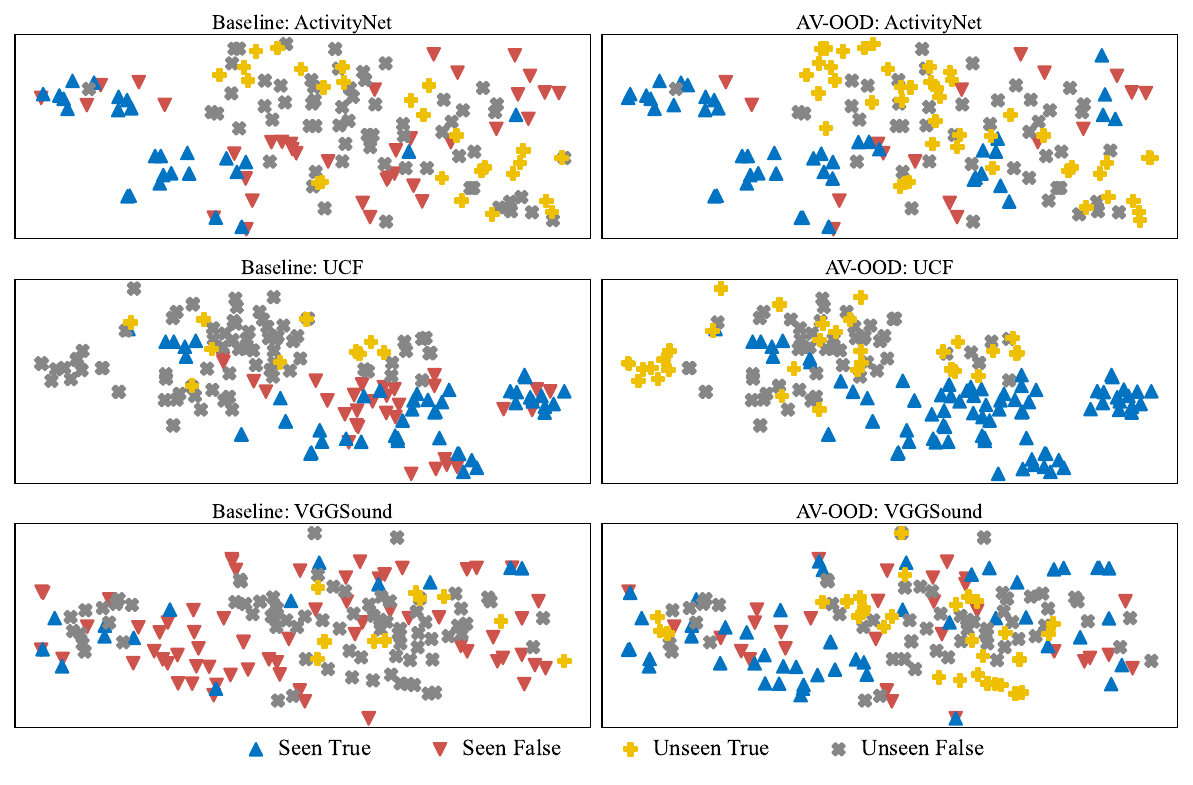}
	\end{center}
	\caption{Comparison between Baseline and our AV-OOD in t-SNE visualizations from test classes across three datasets. In each subplot, we plot 6 seen classes and 6 unseen classes, with 15 samples randomly selected from each category. As to legends, \textit{Ture} and \textit{False} denote correctly or incorrectly classified samples respectively. Thus, as an example, \textit{Seen False} refers to incorrectly classified samples from seen classes. (They can be misclassified as either a seen class or an unseen class. This is not reflected in the figure.)}
	\label{fig:tsne}
\end{figure}
\label{sec:tsne}
In Figure \ref{fig:tsne}, we present t-SNE visualizations for comparison between the \textit{Baseline} AVCA \cite{avca} and our proposed \textit{AV-OOD}. Feature embeddings are obtained using attributes $a \oplus v$, and all of them are grouped into four categories: \textit{Seen True}, \textit{Seen False}, \textit{Unseen True}, and \textit{Unseen False}. 
We can clearly see that \textit{AV-OOD} outperforms \textit{Baseline} in terms of both seen and unseen classes. Although \textit{Baseline} adopts calibrated stacking to reduce bias towards seen classes, this method comes at the cost of sacrificing the accuracy of seen classes. This highlights the superiority of our framework, where the OOD detector and separate expert classifier cooperate, resulting in better overall performance.

\bibliography{supplement}